\documentclass{jltp}
\usepackage{amsmath}
\usepackage{graphics}
\usepackage[sort&compress,comma,super]{natbib}

\newcommand{\br}{\ensuremath{\mathbf{r}}}
\newcommand{\ket}[1]{|#1\rangle}
\newcommand{\bra}[1]{\langle #1|}

\title{Structure and Energetics of Helium Adsorption on Nanosurfaces}

\author{Patrick Huang, Heather D. Whitley, and K. Birgitta Whaley}

\address{Deparment of Chemistry and\\ Kenneth S.\ Pitzer Center for Theoretical Chemistry,\\ University of California, Berkeley, CA 94720, USA}

\runninghead{P.\ Huang, H.~D.\ Whitley, and K.~B.\ Whaley}{Structure and Energetics of Helium Adsorption on Nanosurfaces}

\begin{document}

\maketitle

\begin{abstract}

The ground and excited state properties of small helium clusters,
$^4$He$_N$, containing nanoscale ($\sim$3--10~\AA) planar aromatic
molecules have been studied with quantum Monte Carlo methods.  Ground
state structures and energies are obtained from importance-sampled,
rigid-body diffusion Monte Carlo.  Excited state energies due to
helium vibrational motion are evaluated using the projection operator,
imaginary time spectral evolution technique.  We examine the
adsorption of $N$ helium atoms ($N\leq$~24) on a series of planar
aromatic molecules (benzene, naphthalene, anthracene, tetracene,
phthalocyanine).  The first layer of helium atoms is well-localized on
the molecule surface, and we find well-defined localized excitations
due to in-plane vibrational motion of helium on the molecule surface.
We discuss the implications of these confined excitations for the
molecule spectroscopy.


\end{abstract}

\section{INTRODUCTION}

The utility of He~II as a gentle, quantum matrix for the
high-resolution spectroscopic study of impurity molecules is now
well-established.\cite{stienkemeier01} Accompanying these recent
developments in experimental methodology is an interest in examining
the spectroscopy of large, planar aromatic molecules (PAM) in helium.
These PAMs can be viewed as nanoscale precursors to bulk graphite
surfaces, whose size and geometry can be systematically tuned.  The
experiments have examined the electronic spectra of the molecules, and
can be divided into two general classes: a) those involving small
numbers of helium atoms ($N\leq 17$) around the impurity molecule, and
b) those where the PAM is embedded in a large helium droplet
($N\sim$~10000).

The electronic spectra of PAMs in helium exhibit features not found
for the corresponding bare molecules.  These new features are
presumably due to the presence of helium on the molecule surface, but
a more specific understanding of their origins does not yet exist.  In
the absorption experiments, sharp excitations have been observed at
$\sim 3-30$~K above the electronic origin, which have been
qualitatively attributed to the localized vibrations of helium atoms
adsorbed on the molecule surface.\cite{hartmann98,even01,hartmann02}
This energy range also encompasses the band of states (phonon wings)
associated with collective compressional excitations of the cluster
interior,\cite{hartmann96} and in the large cluster experiments these
localized excitations appear as additional structure superimposed on
the phonon wing sidebands.

For the electronic spectra of a number of different PAMs (tetracene,
Mg-phthalocyanine, and various indole derivatives) in large He
droplets, the peak associated with the electronic origin (zero-phonon
line) also exhibits additional structure, with splittings on the order
of $\sim 1.5$~K.\cite{hartmann01,lindinger01} These splittings are not
observed in the small cluster experiments,\cite{even01} and their
physical origins are not completely understood at this point.  One
possibility is that they are due to the excitation of low-energy modes
involving the collective motion of helium atoms on the molecule
surface.  Theoretical calculations for the He$_N$-benzene cluster
suggest that individual helium atoms are strongly localized on the
molecule surface.\cite{kwon01,huang01b} As we discuss further below,
this is a general feature which we find for a number of other PAMs, so
another possible explanation is that these splittings do not arise
from elastic transitions at all, but are instead due to
inhomogeneities in the local helium environment around the molecule.
Recent emission experiments have also revealed additional features not
previously seen in absorption.  For a number of other PAMs
(phthalocyanine, Mg-phthalocyanine, pentacene), when an electronic and
vibrational mode in the impurity molecule are simultaneously excited,
the resulting emission spectrum as the molecule relaxes to the ground
state is similar to that for absorption, except each line is split into 
two peaks that are separated by $\sim14.8$~K.\cite{lehnig03} 
This suggests that the helium dynamics after electronic excitation 
are more complicated than previously thought.

In this work, we present the results of quantum Monte Carlo
calculations for small numbers of helium atoms ($N\leq 24$) around
various PAMs.  The strength and anisotropy of the molecule-helium
interaction gives rise to a strongly localized layer of helium atoms
on the molecule surface in the cluster ground states.  This
localization has important consequences for the excited
states,\cite{huang01b} and we report some preliminary results for
excitations due to the vibrational motion of helium atoms in this
first monolayer.  We propose that the strong ground state localization
implies similar confinement of the excited state, which may
consequently affect the dynamics of energy transfer.

\section{THEORY AND METHODS}

The Hamiltonian for the $^4$He$_N$-PAM cluster consists of the
rigid-body kinetic energy for the PAM, the translational kinetic
energy for the $N$ $^4$He atoms, and the total potential energy
$\hat{V}$.  This $\hat{V}$ is taken as a sum over two-body
helium-helium\cite{aziz87} and helium-PAM interactions, where the
helium-PAM interaction $V_{\mathrm{I}}$ is a sum of atom-atom
Lennard-Jones pair potentials in the principal axis frame of the
molecule:
\begin{equation}
V_{\mathrm{I}}(\br) = \sum_{\alpha} 4\epsilon_{\alpha}\left[\left(\frac{\sigma_{\alpha}}{|\br-\br_{\alpha}|}\right)^{12} - \left(\frac{\sigma_{\alpha}}{|\br-\br_{\alpha}|}\right)^6\right].
\end{equation}
The sum over $\alpha=\mathrm{C,H,N}$ runs over the individual atoms of
the molecule situated at the position $\br_{\alpha}$, and
Lennard-Jones parameters $\epsilon_{\alpha},\sigma_{\alpha}$ were
taken from Vidali {\em et al.}\cite{vidali83} The potentials evaluated
in this manner are qualitatively similar.  Each of the potentials for
the linear PAMs exhibits minima near the center of the aromatic rings.
The global minima in the potential of each molecule occur at $\sim
2.7$~\AA\ above and below the molecule surface.  The potential for
tetracene at this distance from the molecule surface is shown in
Figure~\ref{fig:density}c.

Ground state properties are obtained using diffusion Monte Carlo (DMC)
methods, in particular the importance-sampled rigid body diffusion
Monte Carlo algorithm.\cite{viel01} The trial functions used here are
products of two-body factors,
\begin{equation}
\Psi_{T} = \prod_{j=1}^N e^{-t_{\mathrm{I}}(\br_j)} \prod_{i<j}^N e^{-t_{\mathrm{He}}(|\br_j-\br_i|)} \label{eq:psitrial}
\end{equation}
where $t_{\mathrm{I}}$ and $t_{\mathrm{He}}$ describe helium-PAM and
helium-helium correlations, respectively:
\begin{equation}
t_{\mathrm{I}}(\br) = \sum_{\alpha}\left(\frac{c_{\alpha}}{|\br-\br_{\alpha}|}\right)^5 + ax^2 + by^2 + cz^2,\quad t_{\mathrm{He}}(r) = \left(\frac{c_{\mathrm{He}}}{r}\right)^5.
\end{equation}
The trial function parameters $c_{\alpha},c_{\mathrm{He}},a,b,c$ are
obtained by minimizing the variational energy of $\hat{H}$ with
respect to $\Psi_T$.  Excited state energies are obtained using the
projection operator, imaginary time spectral evolution (POITSE)
method.\cite{huang01b} This yields the spectral density function
\begin{equation}
\kappa(E) \propto \sum_n |\bra{\phi_n}\hat{A}^{\dagger}\ket{\Psi_T}|^2 \delta(E-E_n+E_0),\quad \hat{A}^{\dagger} = \sum_{j=1}^N \hat{a}^{\dagger}_j \label{eq:kappa},
\end{equation}
where $\{\phi_n\}$ and $\{E_n\}$ are the set of eigenfunctions and
associated eigenvalues of the Hamiltonian $\hat{H}$, respectively.
The many-body operator $\hat{A}^{\dagger}$ is a Bose-symmetrized sum
of one-body excitation operators $\hat{a}^{\dagger}$, and is chosen to
connect the ground state $\Psi_T$ to the excited state(s) of interest.

\section{RESULTS AND DISCUSSION}

\begin{figure}
\begin{center}
\resizebox{0.9\textwidth}{!}{\includegraphics{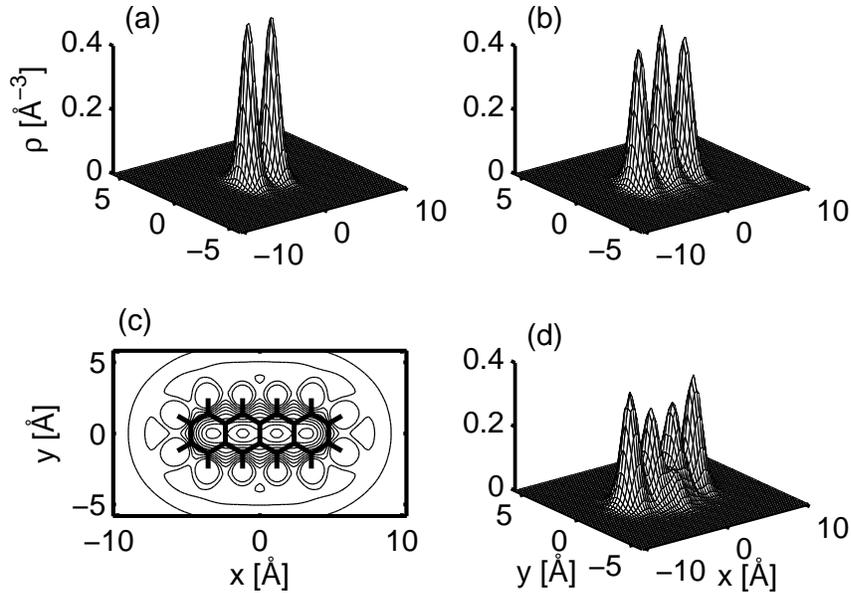}}
\end{center}
\caption{\label{fig:density}Helium density distributions at a distance
of $z=2.9$~\AA\ above the molecule plane for (a)
$^4$He$_4$-naphtalene, (b) $^4$He$_6$-anthracene, and (d)
$^4$He$_8$-tetracene.  A cut of the tetracene potential at
$z=2.7$~\AA\ is also shown in (c), where contour lines run from
$-160$~K to 0~K, in increments in $15$~K.  The tetracene molecule on
the $z=0$ plane is drawn in bold lines.}
\end{figure}

\begin{figure}
\begin{center}
\resizebox{0.9\textwidth}{!}{\includegraphics{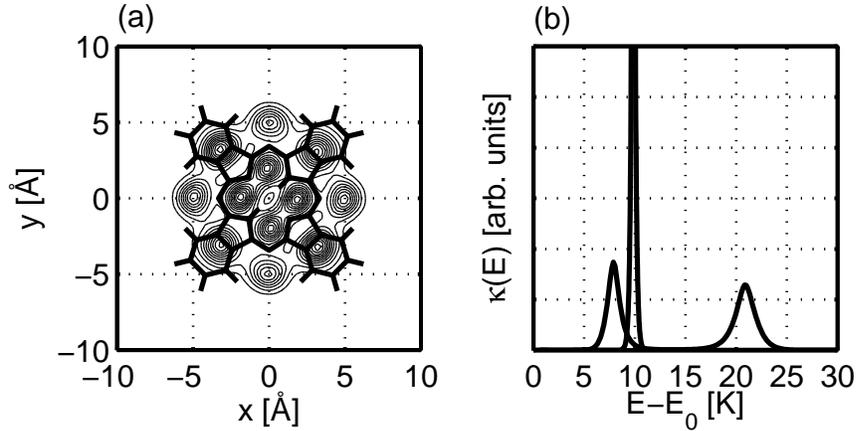}}
\end{center}
\caption{\label{fig:excited}Left: Helium density distribution at a
distance of $z=3.2$~\AA\ above the molecule plane for
$^4$He$_{24}$-phthalocyanine.  The phthalocyanine molecule on the
$z=0$ plane is drawn in bold lines.  Right: Spectral density function
for $^4$He$_8$-tetracene.}
\end{figure}

Ground state helium density distributions for $^4$He$_4$-naphthalene,
$^4$He$_6$-anthracene, $^4$He$_8$-tetracene, and
$^4$He$_{24}$-phthalocyanine were obtained using the quantum Monte
Carlo methods described above.  The number of $^4$He atoms around each
of the PAMs studied here corresponds to full coverage of both sides of
the nanosubstrate surface.  In all cases, the global density maxima
occur at a distance of $z\sim 2.9-3.2$~\AA\ above and below the
molecule plane.  Figures~\ref{fig:density}a, \ref{fig:density}b,
\ref{fig:density}d, and \ref{fig:excited}a show cuts of the local
helium density at this distance, along a plane parallel to the
molecule surface.
It is apparent that the local helium densities consist of
well-separated peaks, each corresponding to a single $^4$He atom.  For
the linear PAMs, each individual $^4$He atom is approximately situated
above a six-membered carbon ring.  For phthalocyanine, we find that
this surface can support up to 12 $^4$He atoms on each side.  At full
coverage, this 2D layer of $^4$He on phthalocyanine consists of four
$^4$He atoms grouped near the center of the molecule, surrounded by an
outer ring of eight $^4$He atoms.

Previous work with the $^4$He$_N$-benzene system ($N=1,2,3,14$) has
revealed similar ground state features.\cite{huang01b} The benzene
system can be viewed as the simplest $^4$He$_N$-PAM system, serving as
a building block for models of more complicated one- and
two-dimensional nanosubstrates.  In that study, not only were the
ground states observed to be strongly localized, but the {\em excited
states} also exhibited localized character.  That is, excitations due
to the localized vibrational motion of $^4$He on the benzene surface
were found, with energies of up to $\sim 23$~K above the ground state.
We propose here that these excited state features exist generally for
larger $^4$He-PAM systems, and that they are a result of the strong
localization of the helium ground state density distribution.
Figure~\ref{fig:excited}b shows the spectral density function
[Eq.~(\ref{eq:kappa})] for $^4$He$_8$-tetracene, evaluated using the
POITSE methodology discussed above.
For a choice of $\hat{a}^{\dagger}=x$ and $x^2$, where the
$x$-direction is taken to be along the long axis of the molecule, we
find excitations at $6.3,9.9,20.8$~K above the ground state.  These
states correspond to the collective vibrational motion of the $^4$He
atoms moving on the molecule surface along the $x$-direction.  Work is
currently in progress to extend these calculations to larger $N$, in
order to ascertain whether the collective character of these
excitations persists as additional atoms are added around the
molecule.

A helium excitation confined on the molecule surface would have
important consequences for spectroscopy, particularly for the
dynamics of energy dissipation. Typically, 
the molecule is excited from the electronic ground state
$S_0$ to an electronic excited state $S_1$.  For a weakly interacting
small molecule in a vibrationally excited state of $S_1$, this excess
vibrational energy is transferred to the local helium environment.  It
is generally assumed that this energy is rapidly carried away from the
molecule due to the large thermal conductivity of
He~II.\cite{tilley86} In a helium droplet, this energy is dissipated
as helium atoms boil off the surface of the droplet, and so the
relaxation back to $S_0$ derives primarily from the ground vibrational
state of $S_1$.  On the other hand, if the excess vibrational energy
of the molecule is transferred to the helium and trapped near the
surface of the molecule, the resulting local helium environment could
be very different from that of the ground vibrational state and 
would then correspond 
to a metastable state.  
These two situations could give rise to two
distinct $S_1$ surfaces, resulting in a split
 emission spectrum similar to that observed by 
Lehnig and Slenczka.~\cite{lehnig03}  Calculation of the local helium 
density for the $S_1$ state, which requires a currently unknown interaction 
potential, 
would facilitate a more detailed analysis.


\begin{thebibliography}{10}

\bibitem{stienkemeier01}
F. Stienkemeier and A.~F. Vilesov, {\it J.\ Chem.\ Phys.} {\bf 115},  10119  (2001).

\bibitem{hartmann98}
M. Hartmann, A. Lindinger, J.~P. Toennies, and A.~F. Vilesov, {\it Chem.\ Phys.} {\bf
  239},  139  (1998).

\bibitem{even01}
U. Even, I. Al-Hroub, and J. Jortner, {\it J.\ Chem.\ Phys.} {\bf 115},  2069
  (2001).

\bibitem{hartmann02}
M. Hartmann, A. Lindinger, J.~P. Toennies, and A.~F. Vilesov, 
{\it Phys.\ Chem.\ Chem.\ Phys.} {\bf 4},  4839  (2002).

\bibitem{hartmann96}
M. Hartmann {\it et~al.}, {\it Phys.\ Rev.\ Lett.} {\bf 76},  4560  (1996).

\bibitem{hartmann01}
M. Hartmann, A. Lindinger, J.~P. Toennies, and A.~F. Vilesov, {\it J.\ Phys.\ Chem.\ A} {\bf 105},  6369  (2001).

\bibitem{lindinger01}
A. Lindinger, E. Lugovoj, J.~P. Toennies, and A.~F. Vilesov, {\it Z.\ Phys.\ Chem.} {\bf 215},  401  (2001).

\bibitem{kwon01}
Y. Kwon and K.~B. Whaley, {\it J.\ Chem.\ Phys.} {\bf 114},  3163  (2001).

\bibitem{huang01b}
P. Huang and K.~B. Whaley, {\it Phys.\ Rev.\ B} {\bf 67},  155419  (2003).

\bibitem{lehnig03}
R. Lehnig and A. Slenczka, {\it J.\ Chem.\ Phys.} {\bf 118},  8256  (2003).

\bibitem{aziz87}
R.~A. Aziz, F.~R.~W. McCourt, and C.~C.~K. Wong, {\it Mol.\ Phys.} {\bf 61},  1487
  (1987).

\bibitem{vidali83}
G. Vidali, M.~W. Cole, W.~H. Weinberg, and W.~A. Steele, {\it Phys.\ Rev.\ Lett.} {\bf 51},  118  (1983).

\bibitem{viel01}
A. Viel, M.~V. Patel, P. Niyaz, and K.~B. Whaley, {\it Comp.\ Phys.\ Com.} {\bf 145}, 24  (2002).

\bibitem{tilley86}
D.~R. Tilley and J. Tilley, {\em Superfluidity and Superconductivity} (Hilger, Bristol, UK, 1986).

\end{thebibliography}
\end{document}